\documentclass[letterpaper]{jpconf}
\usepackage{graphicx}
\begin{document}

\title{Calculation of the nucleon axial charge in lattice QCD}

\author{D.B.~Renner$^1$, R.G.~Edwards$^2$, G.~Fleming$^3$, Ph.~H\"agler$^4$, J.W.~Negele$^5$,
K.~Orginos$^{2,6}$, A.V.~Pochinsky$^5$, D.G.~Richards$^2$ and W.~Schroers$^7$}

\address{$^1$ University of Arizona, Department of Physics, 1118 E 4th St, Tucson AZ 85721}
\address{$^2$ Thomas Jefferson National Accelerator Facility, Newport News, VA 23606}
\address{$^3$ Sloane Physics Laboratory, Yale University, New Haven, CT 06520}
\address{$^4$ Institut f\"ur Theoretische Physik, TU M\"unchen, D-85747 Garching, Germany}
\address{$^5$ Center for Theoretical Physics, Massachusetts Institute of Technology, Cambridge, MA 02139}
\address{$^6$ Department of Physics, College of William and Mary, Williamsburg VA 23187}
\address{$^7$ John von Neumann-Institut f\"ur Computing NIC/DESY, D-15738 Zeuthen, Germany}

\begin{abstract}
Protons and neutrons have a rich structure in terms of their constituents,
the quarks and gluons.  Understanding this structure requires solving
Quantum Chromodynamics (QCD).  However QCD is extremely complicated,
so we must numerically solve the equations of QCD using a method
known as lattice QCD.  Here we describe a typical lattice QCD calculation
by examining our recent computation of the nucleon axial charge.
\end{abstract}

\section{Introduction}
Quantum Chromodynamics is the theory of the Strong Force of
nature.  The theory dictates how quarks, the fundamental 
components of ordinary matter, interact by exchanging particles 
known as gluons.  The interactions between quarks and gluons
are sufficiently strong to bind them into bound states such as 
protons, neutrons, and pions.  This phenomenon not only gives 
these bound states a rich and detailed structure but also makes 
pencil-and-paper calculations of many properties of QCD impossible.
Consequently we numerically solve the equations of QCD by using a 
discrete space-time lattice.  We use this method, known as lattice 
QCD, to calculate the properties of protons and neutrons, 
collectively known as nucleons, and to ask how nature constructs 
such objects from their quark and gluon constituents.  Here we 
describe a typical calculation, the calculation of the nucleon 
axial charge, $g_A$, to illustrate our effort to determine the 
quark and gluon structure of the nucleon.

\subsection{Quantum Chromodynamics}
\begin{figure}[htb]
\begin{minipage}{18pc}
\includegraphics[width=18pc]{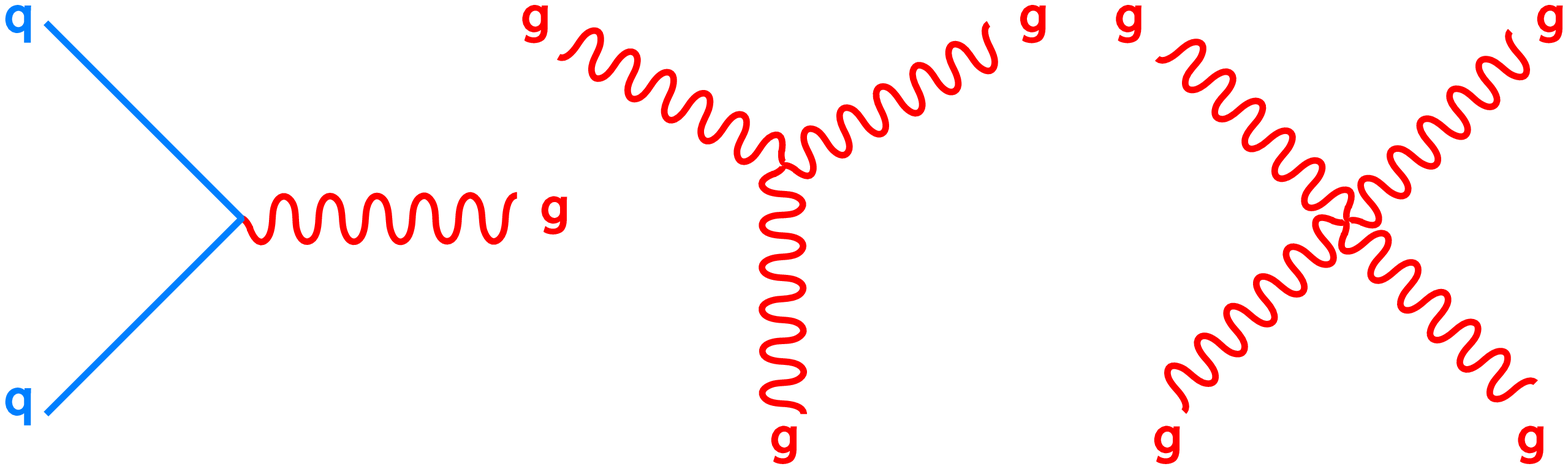}
\caption{\label{qcd_feynman_rules}Fundamental quark and gluon interactions of QCD}
\end{minipage}
\hspace{1.5pc}
\begin{minipage}{18pc}
\includegraphics[width=16.25pc]{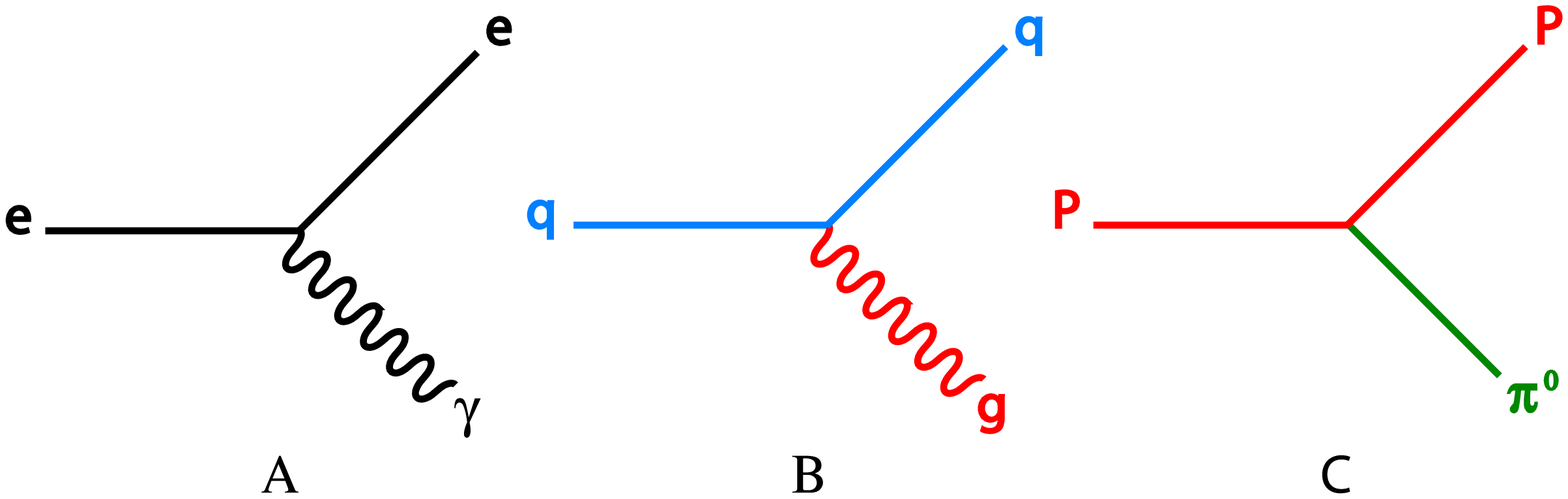}
\caption{\label{axial_charge}Examples of charges in QED (A), QCD (B) and $\chi$PT (C)}
\end{minipage}
\end{figure}

Quantum Chromodynamics is the theory within the Standard Model of 
particle physics that explains the Strong Force between the 
quarks and gluons.  There are six quarks named up, down, strange, 
charm, bottom, and top.  The fundamental interactions among the 
quarks and gluons are conceptually quite simple 
but often very complicated in practice.
In Figure~\ref{qcd_feynman_rules} we show the three fundamental interactions between quarks
and gluons in QCD.  All processes in QCD are built by combining
these three basic interactions in all possible ways (Figure~\ref{nucleon_propagator}).

\subsection{Nucleon Axial Charge} 
There are many kinds of charges in particle
physics.  Electrons have an electric charge to which photons couple (Figure~\ref{axial_charge}-A).  Quarks
have a color charge to which gluons couple (Figure~\ref{axial_charge}-B).  Protons have an axial charge 
to which pions couple (Figure~\ref{axial_charge}-C).  The electric and color charges are fundamental properties
of elementary particles and have prescribed values, however the axial charge of the 
proton is a consequence of its quark and gluon structure and hence
must be measured experimentally or calculated theoretically.

\section{Nucleon Axial Charge}
The nucleon axial charge, $g_A$, is a particularly illuminating property of the
nucleon.  It reveals how the up and down quark intrinsic spin contribute to the
spin of the proton and neutron. %(Figure~\ref{proton_spin}).  
It is a basic ingredient in the expressions that
determine how the neutron decays. %(Figure~\ref{beta_decay}).  
It is a measure of the extent to which 
spontaneous breaking of chiral symmetry impacts the structure of the proton and
neutron. %(Figure~\ref{chiral_condensate}).  
And it determines how nucleon properties vary as you change the 
quark masses in QCD. %(Figure~\ref{chipt}).}
The current experimental measurement is $g_A=1.2695\pm0.0029$~\cite{Eidelman:2004wy}.
Our lattice QCD calculation gives $g_A=1.212\pm0.084$~\cite{Edwards:2005ym} 
which has a $7\%$ fractional error and agrees
with the experimental determination.

\subsection{Proton Spin Physics}
\begin{figure}[htb]
\begin{minipage}{18pc}
\begin{center}
\includegraphics[width=9.5pc]{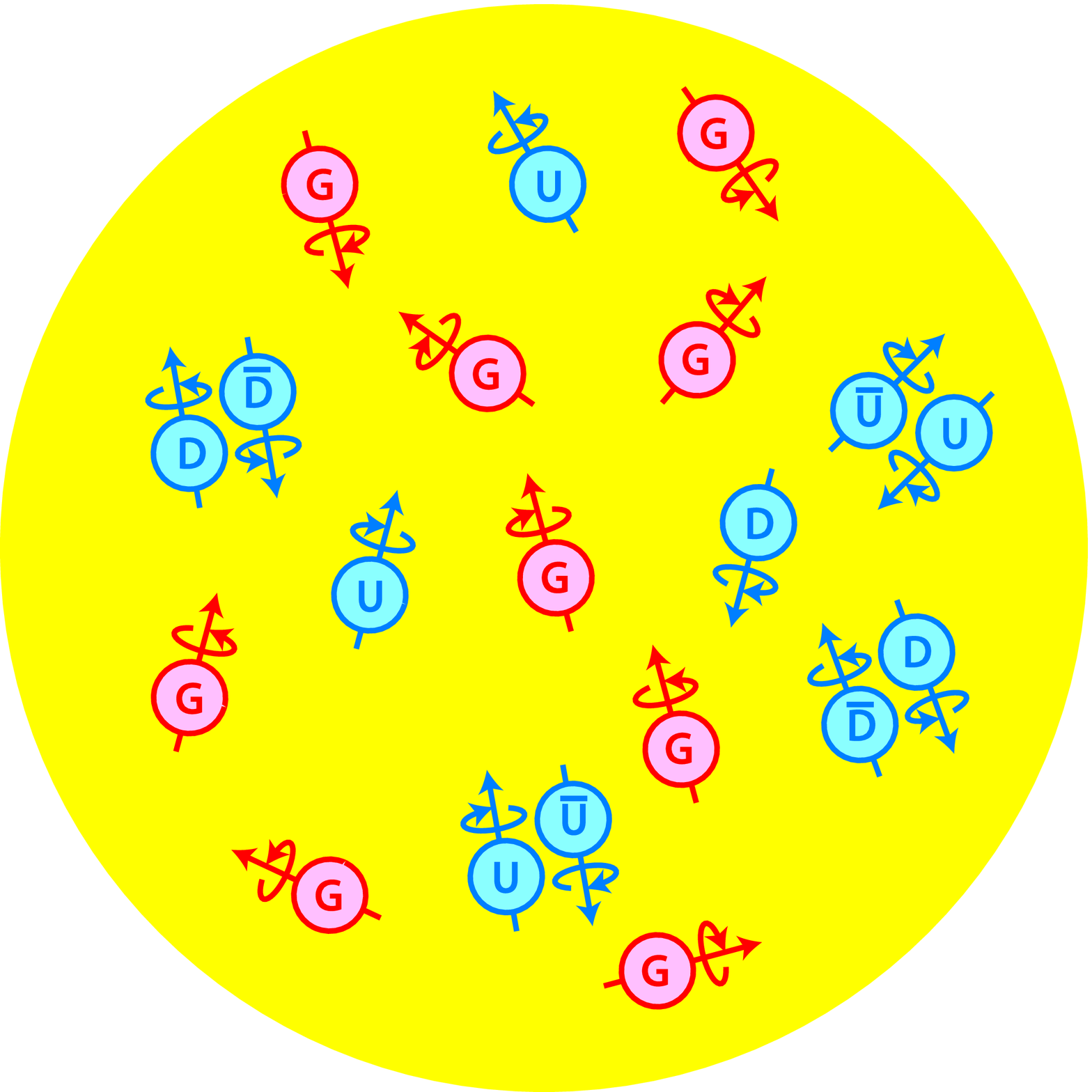}
\end{center}
\caption{\label{proton_spin}
The nucleon is a composite object whose
spin arises from the combination of quark and gluon spin
as well as quark and gluon orbital motion.
%The nucleon is a composite object whose
%intrinsic angular momentum, known as spin, arises from the combination of quark and gluon spin
%as well as quark and gluon orbital angular momentum.  The axial charge determines
%the difference of the up quark and down quark spin contributions.
}
\end{minipage}
\hspace{1.5pc}
\begin{minipage}{18pc}
\includegraphics[width=18pc]{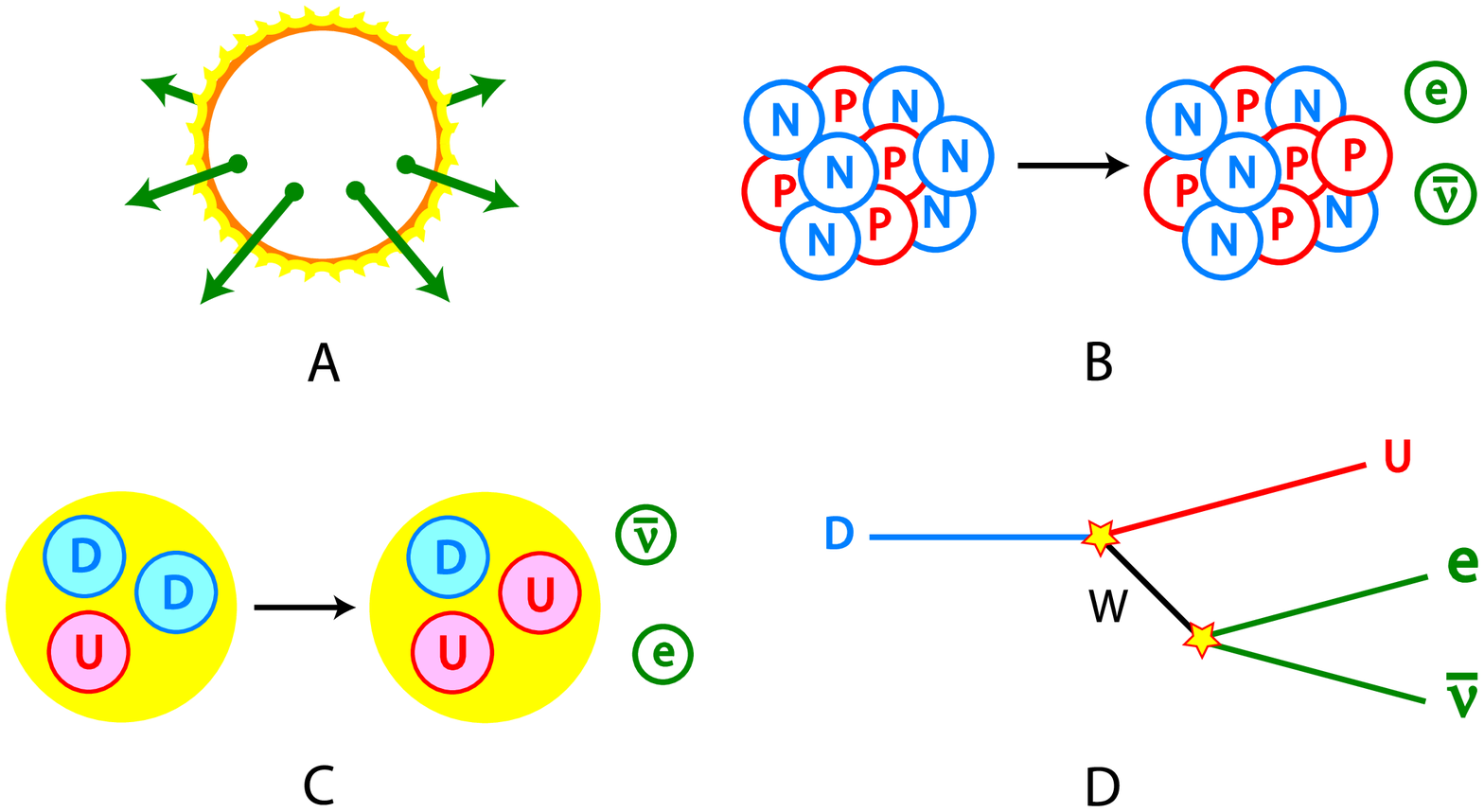}
\caption{\label{beta_decay}
These pictures illustrate the
relevance of the nucleon axial charge to astrophysics (A), nuclear
physics (B), hadronic physics (C) and particle physics (D).
}
\end{minipage}
\end{figure}

Both elementary and composite particles have
intrinsic angular momentum known as spin.  The
rotational properties of three dimensional space dictate
that the proton must have precisely a spin 
of $1/2$. %(in units of $\hbar$).  
Additionally the 
constituents of the proton, quarks and gluons, must 
have exactly a spin of $1/2$ and $1$ respectively.
However the spin of the quarks and gluons may be orientated in any direction.
Furthermore they may also move within the proton giving rise to
additional orbital angular momentum.  Thus the proton spin
has four sources:\ intrinsic quark and gluon spin as well
as quark and gluon orbital angular momentum (Figure~\ref{proton_spin}), and these 
four contributions must somehow conspire to sum to precisely
$1/2$.  Of these four contributions, the axial charge determines
the difference of the intrinsic up and down quark 
contributions to the nucleon spin: $g_A=(0.75)-(-0.52)=1.2695\pm 0.0029$~\cite{Eidelman:2004wy,Hannappel:2005vp}.

\subsection{Neutron Beta Decay}
There is no experimental evidence that the proton ever decays.
However
the neutron does decay, and it does so with a mean lifetime of nearly
15 minutes.
The mechanism underlying neutron decay is the conversion of
a down quark inside of a neutron into an up quark.  This process is
mediated by the Weak Force and is a fundamental interaction in the 
Standard Model.  The result is
that a neutron becomes a proton while emitting two other particles: an electron and electron 
anti-neutrino.  The rate of neutron decay is determined by several 
parameters of the Standard Model as well as the nucleon axial charge.

Neutron beta decay impacts a range of physical phenomena from 
astrophysics, to nuclear physics, to hadronic physics, and down to
particle physics.
In astrophysics, beta decay of the neutron, 
$\mathrm{n}\rightarrow\mathrm{p}\,\mathrm{e}\,\bar{\mathrm{\nu}}$,
and electron capture on the proton,
$\mathrm{e}\,\mathrm{p}\rightarrow\mathrm{n}\,\mathrm{\nu}$,
play an important role in the emission of neutrinos from stars (Figure~\ref{beta_decay}-A).
In nuclear physics, nuclear beta decay occurs when a neutron in 
the nucleus undergoes beta decay to a proton (Figure~\ref{beta_decay}-B).
In hadronic physics, 
the axial charge can be determined from the beta decay of the neutron (Figure~\ref{beta_decay}-C).
And in particle 
physics, the axial form factor of the nucleon can be measured in
neutrino-nucleon scattering which probes the weak interactions of the
up and down quarks (Figure~\ref{beta_decay}-D).

\subsection{Spontaneous Chiral Symmetry Breaking}
\begin{figure}[htb]
\begin{minipage}{18pc}
\includegraphics[width=18pc]{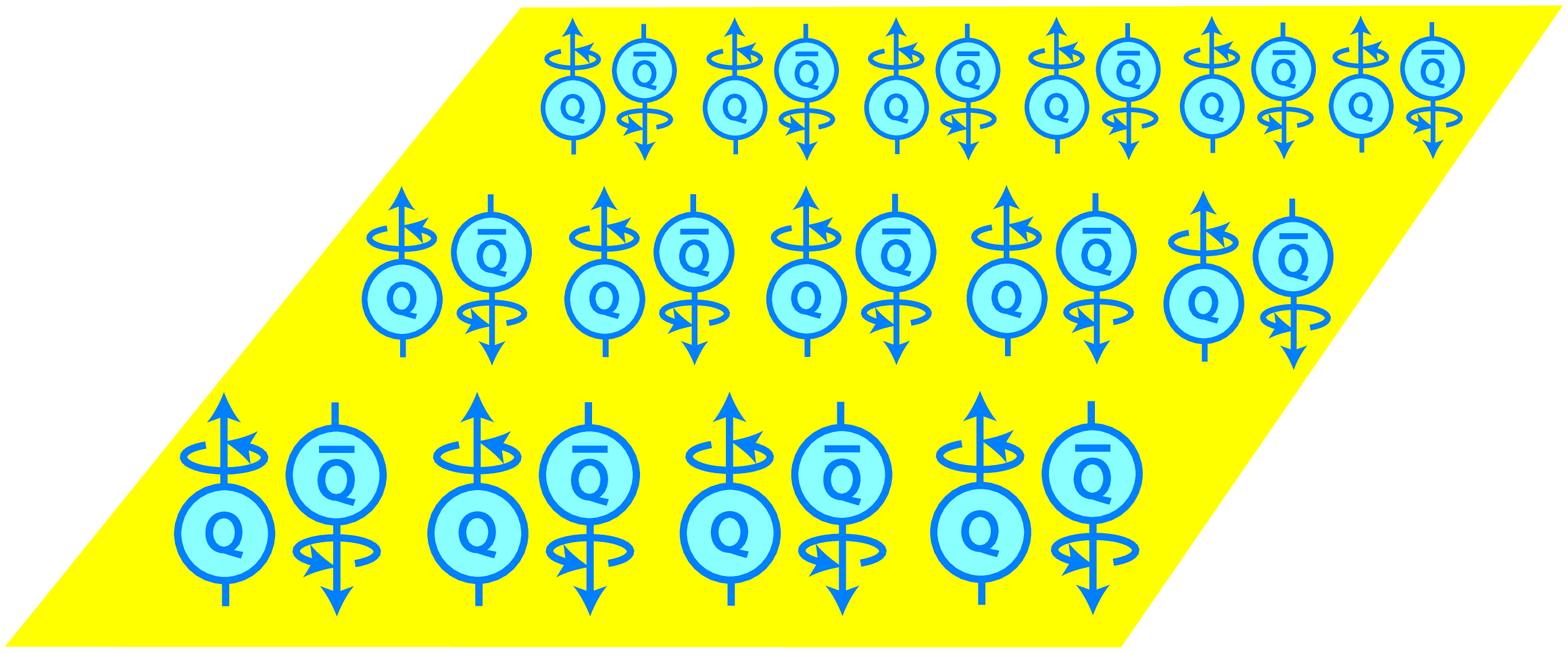}
\caption{\label{chiral_condensate}
The QCD vacuum is filled with
a condensate of quark-anti-quark pairs 
that gives the axial charge a non-zero value.
%The QCD vacuum is filled with
%a condensate of quark-anti-quark pairs, 
%right-handed quarks and left-handed anti-quarks
%or left-handed quarks and right-handed anti-quarks,
%that break chiral symmetry, a fundamental symmetry of the interactions
%of massless quarks and gluons, and gives the axial charge a non-zero value.
}
\end{minipage}
\hspace{1.5pc}
\begin{minipage}{18pc}
\includegraphics[width=18pc]{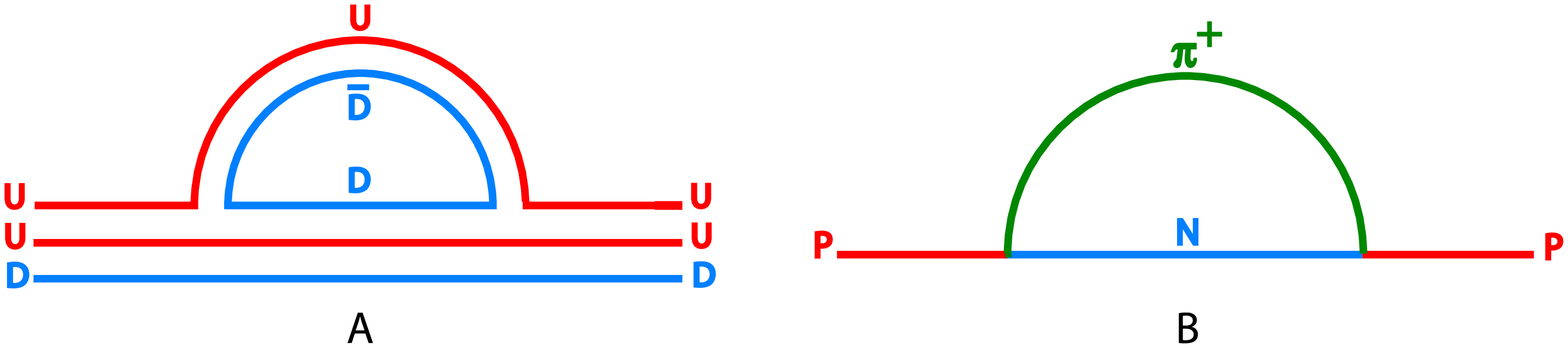}
\caption{\label{chipt}
Interactions in QCD involving quarks and gluons (A) are
replaced by equivalent interactions of pions and nucleons (B) in chiral perturbation theory.  
The axial charge determines the strength of these pion and nucleon interactions.
}
\end{minipage}
\end{figure}

Chiral symmetry is a fundamental symmetry of the quark and
gluon interactions in the limit of zero quark masses.  Quarks with their spin aligned to
their direction of motion, called right-handed, have gluon interactions identical to 
quarks with their spin anti-aligned, called left-handed.  However the
vacuum of QCD breaks this symmetry.  Pairs of
right-handed quarks and left-handed anti-quarks or left-handed quarks 
and right-handed anti-quarks condense and fill
the vacuum with quark-anti-quark pairs (Figure~\ref{chiral_condensate}) that break chiral symmetry.
%An elementary argument demonstrates that the 
%axial charge of the nucleon would vanish in the
%limit of zero quark masses if this symmetry breaking did not occur, hence
%the non-zero value for $g_A$ indicates the extent to which
%spontaneous chiral symmetry breaking impacts the physics of protons
%and neutrons.
An elementary argument demonstrates that the 
axial charge of the nucleon would vanish in the
limit of zero quark masses if this symmetry breaking did not occur, indicating
the intimate connection between $g_A$ and the spontaneous breaking of chiral symmetry in QCD.

\subsection{Chiral Perturbation Theory}
Chiral perturbation theory
describes the physics of QCD with light quark masses.  It does so by
replacing the quark and gluon interactions
by a set of pion and nucleon interactions (Figure~\ref{chipt})
chosen in precisely the correct way to
reproduce the low energy predictions of QCD.
Chiral perturbation theory is particularly useful because it predicts the quark mass dependence
of lattice QCD calculations (Figure~\ref{chiral_ga}).
In particular
the axial charge effects how rapidly the properties of the nucleon vary with
the quark masses.

\section{Lattice QCD}
Numerical calculations are performed in lattice QCD by replacing 
continuous space-time with a discrete lattice.
In this way lattice calculations reduce to the evaluation of very high dimensional integrals.
State-of-the-art lattice QCD
computations use lattices as large as $48^3\times 144$ with $4\times 8$ degrees
of freedom per lattice point which results in just slightly more than half a billion integration variables.  
Such
enormous integrals are evaluated using Monte Carlo methods.  The dominant
limitation of this method is the time required to invert large, poorly conditioned
matrices that account for the effects of quark-anti-quark pairs in the vacuum.  This 
problem becomes increasingly more severe with lighter quark masses, hence
lattice calculations are performed with quark masses heavier than
those in nature and the results are extrapolated to the physical point.

\begin{figure}[htb]
\begin{minipage}{18pc}
\includegraphics[width=18pc]{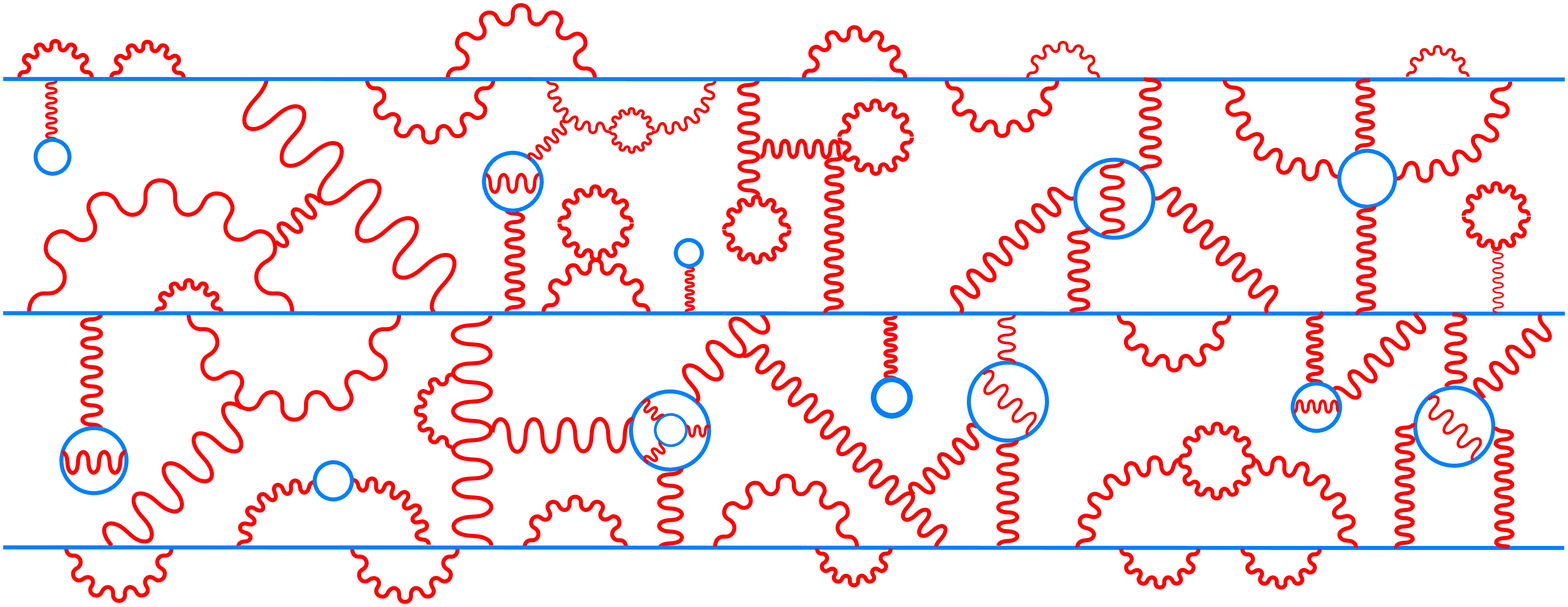}
\caption{\label{nucleon_propagator}
%This is one of infinitely many quark (blue lines) 
%and gluon (red lines) contributions to the propagation of a nucleon.
A nucleon consists of three net quarks
which persist in time from the left to 
the right but are accompanied by arbitrarily many gluons and quark-anti-quark
pairs.
}
\end{minipage}
\hspace{1.5pc}
\begin{minipage}{18pc}
\includegraphics[width=17.5pc,height=6.75pc]{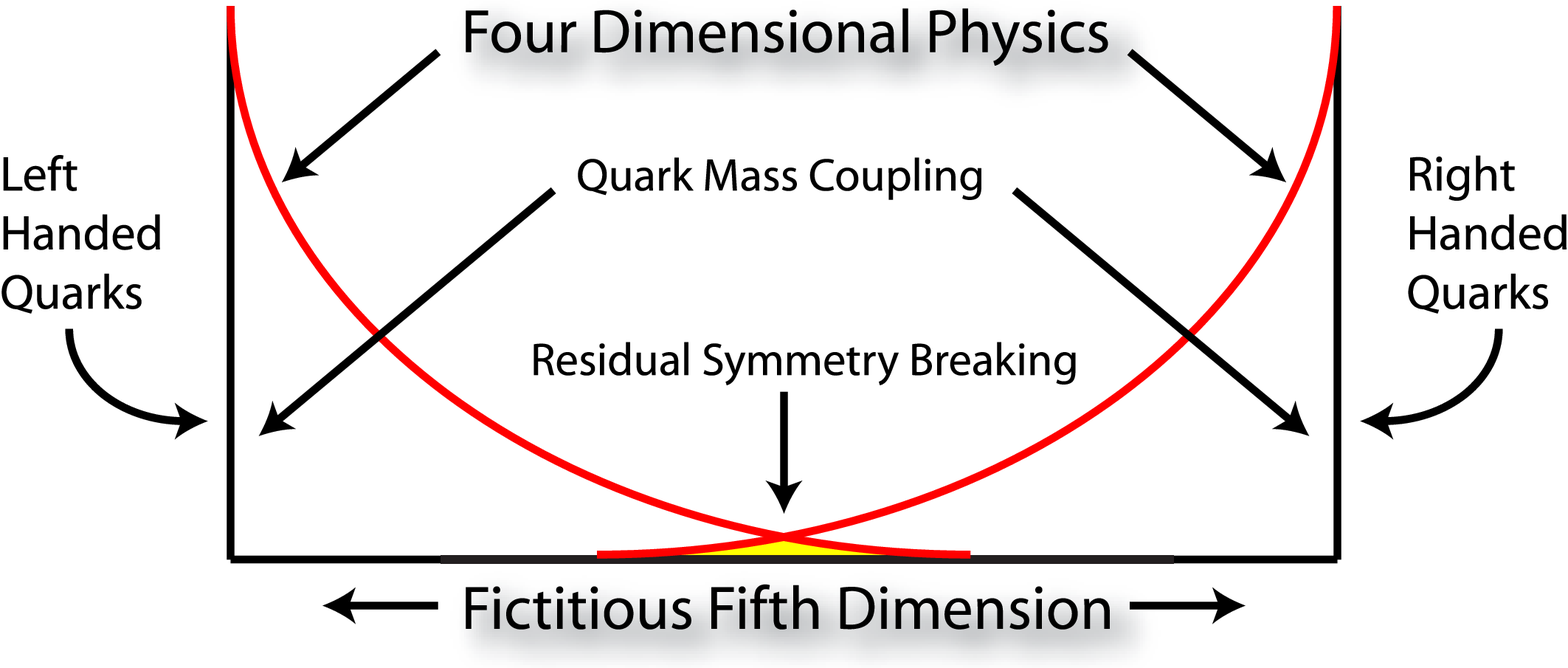}
\caption{\label{domain_wall}
Domain wall fermions use a fictitious fifth dimension to construct 
a formulation of quarks in lattice QCD with
an exact chiral symmetry even on the lattice.
}
\end{minipage}
\end{figure}

\subsection{Domain Wall Fermions}
There are many ways to place quarks on a space-time lattice.
In fact, a broad class of lattice theories all give rise to the same QCD predictions
in the limit of vanishing lattice spacing.  However
there is a ``no-go'' theorem
stating under general assumptions that no \emph{simple} four dimensional 
representation of quarks can have the correct chiral symmetries of QCD.
Domain wall fermions were
invented
to elude this obstacle by using
a fictitious fifth dimension (Figure~\ref{domain_wall}).  The four dimensional physics is exponentially
bound to the edges of the fifth dimension.  All artifacts of the
five dimensional theory vanish and an exact chiral symmetry emerges 
as the extent of the fifth dimension is made larger and larger.

\subsection{Our Latest Lattice Results}
\begin{figure}[htb]
\begin{minipage}{18pc}
\includegraphics[width=18pc]{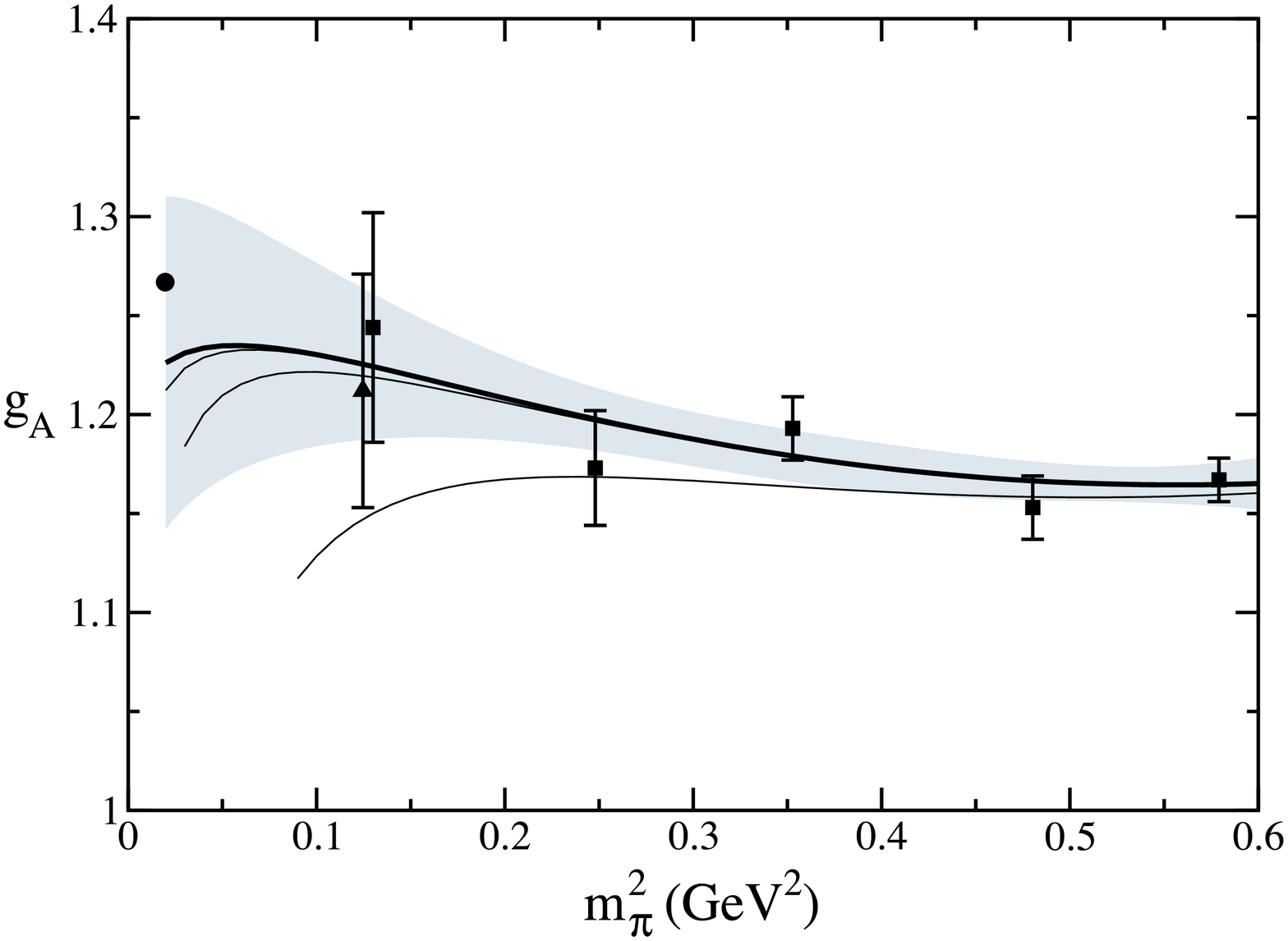}
\caption{\label{chiral_ga}
Our recent lattice QCD results for $g_A$~\cite{Edwards:2005ym}
%The squares and triangle are the results of our lattice
%computation of $g_A$ at each of five pion masses with
%box sizes of $2.5~\mathrm{fm}$ (squares) and $3.5~\mathrm{fm}$ (triangle).  The
%circle is the experimental measurement of $g_A$ at the
%physical pion mass.  The heavy curve and error band are the chiral perturbation
%theory prediction for the pion mass dependence of the axial charge for
%an infinite box size.  The three lighter curves are the chiral perturbation
%theory predictions of the pion mass dependence for box sizes of $3.5~\mathrm{fm}$,
%$2.5~\mathrm{fm}$, and $1.5~\mathrm{fm}$.
}
\end{minipage}
\hspace{1.5pc}
\begin{minipage}{18pc}
\includegraphics[width=17.75pc]{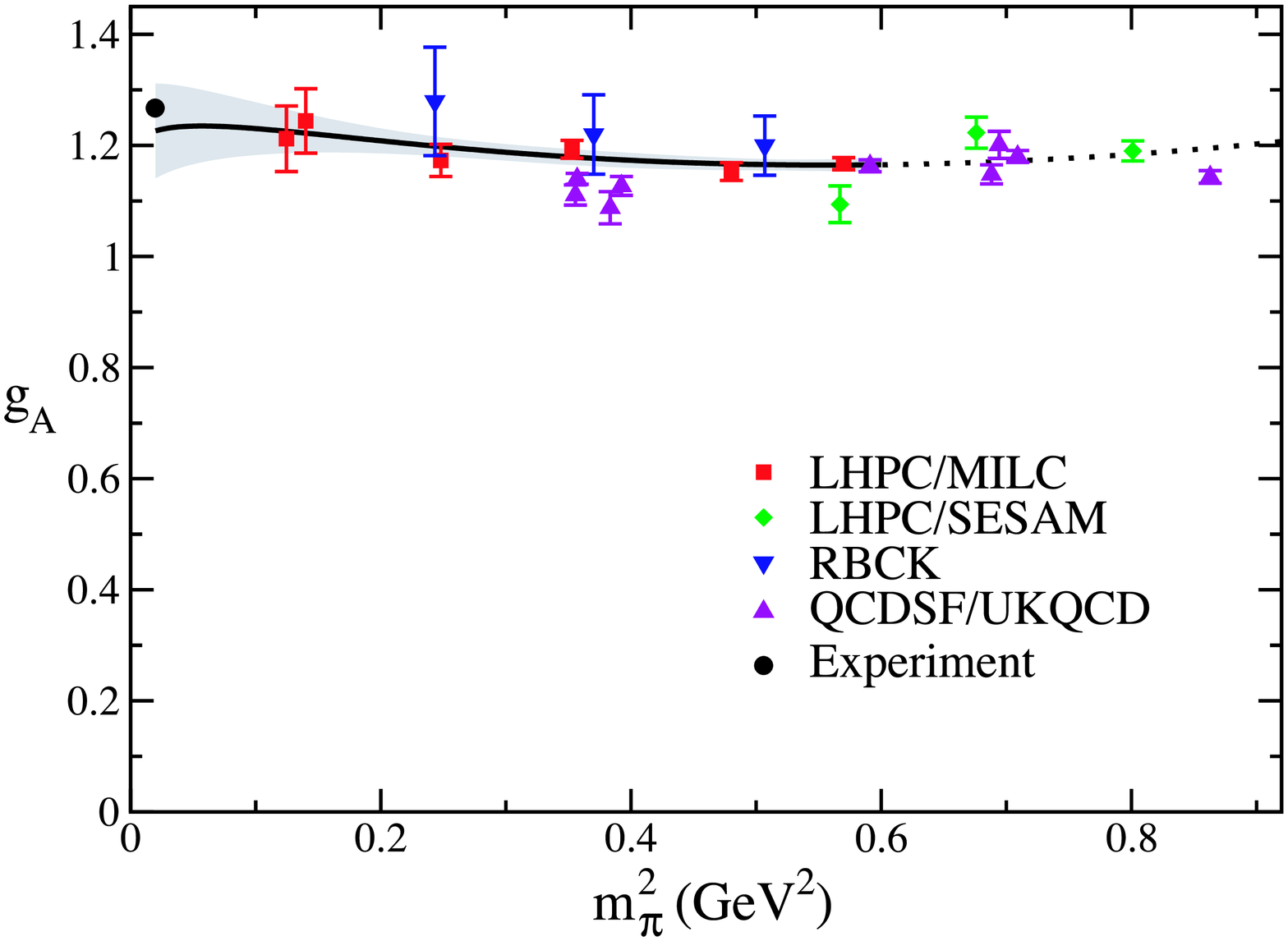}
\caption{\label{world_ga}
World's collection of lattice QCD results for $g_A$~\cite{Edwards:2005ym,Ohta:2004mg,Khan:2004vw,Dolgov:2002zm}
%The results here are the world's collection of lattice calculations
%of $g_A$.  The red points are our calculation
%from Figure~\ref{chiral_ga}.  The green points are the results from our previous
%calculation [5,6], and the blue and purple points are the results from [7] and [8] respectively.
}
\end{minipage}
\end{figure}

We use the gluon configurations from the MILC
collaboration~\cite{Bernard:2001av}
for the evaluation of the lattice QCD integrals.  Additionally we use domain wall
fermions to represent the quarks inside the nucleon.  Our current calculation is done
at five quark masses, equivalently pion masses, at one volume
and additionally we have repeated the lightest calculation at a larger
volume to check for finite size errors (Figure~\ref{chiral_ga}).  We then match the lattice results 
to chiral perturbation theory in order to extrapolate to both the zero quark mass
and infinite volume limits.  The final result for our computation of the axial charge
is shown in Figure~\ref{chiral_ga} for which the lattice computation of 
$g_A=1.212\pm0.084$~\cite{Edwards:2005ym} agrees to within the errors with the experimental determination of
$g_A=1.2695\pm0.0029$~\cite{Eidelman:2004wy}.

\subsection{World's Lattice Results}
There have been several lattice QCD calculations of the nucleon axial charge,
however, only four calculations in the world have included the effects of quark-anti-quark 
pairs in the vacuum.  We performed the first such calculation~\cite{Dolgov:2002zm}, and we 
have since extended the calculation as described here~\cite{Edwards:2005ym}.
The dominant source 
of error in these calculations is the extrapolation to the physical quark 
masses, 
and as Figure~\ref{world_ga} shows, our method has allowed us to calculate
the axial charge with lighter quark masses than any other group.

\section{Conclusions}
The calculation of the nucleon axial charge described here 
represents a significant advance in our ability to compute properties
of the proton and neutron from lattice QCD.
Current 
calculations are focused on reducing the error at the lightest
quark masses in Figure~\ref{chiral_ga}, extending our calculations to
lighter masses still, and additionally calculating with a smaller lattice
spacing.  Continued improvement in our calculation of the axial charge
will bolster the strength behind our computation of nucleon observables
that can not be measured experimentally but yet reveal a great deal
about the quark and gluon structure of the nucleon.

\section*{References}
\bibliography{ga_proc}

\end{document}